\documentclass[reqno]{amsart}
\usepackage[active]{srcltx}
\usepackage{deleq}  
\usepackage{epsfig}
\usepackage{amssymb}
\usepackage{amsfonts}
 \usepackage{latexsym}
\usepackage{amsmath}
\usepackage{amsfonts} \usepackage{amsthm}

\newtheorem{Proposition}{Proposition}
  \newtheorem{Remark}[Proposition]{Remark}
  
  \newtheorem{Lemma}[Proposition]{Lemma}
  
  \newtheorem{Theorem}[Proposition]{Theorem}

\renewcommand{\theequation}{\thesection.\arabic{equation}}
  \makeatletter
\@addtoreset{equation}{section}
\makeatother 
       \def\z{\noindent}

    \def\z{\noindent}  
 \def\Box{{\hfill\hbox{\enspace${\sqre}$}} \smallskip}
    \def\sqr#1#2{{\vcenter{\vbox{\hrule height .#2pt
                             \hbox{\vrule width .#2pt height#1pt \kern#1pt
                                   \vrule width .#2pt}
                             \hrule height .#2pt}}}}
 \def\sqre{\mathchoice\sqr54\sqr54\sqr{4.1}3\sqr{3.5}3}

 \def\bchi{\mbox{\raisebox{.4ex}{\begin{large}$\chi$\end{large}}}}
     \def\CC{\mathbb{C}}
    
    \def\NN{\mathbb{N}}
    
    \def\RR{\mathbb{R}}
    \def\ZZ{\mathbb{Z}}

\begin{document}

\author{O. Costin, R. D. Costin, and J. L.  Lebowitz$^1$ {$_{\mbox{
        Department of Mathematics, Rutgers University}}$}}
\thanks{$^1$Also Department of Physics.}  \title{{Time asymptotics of
    the Schr\"odinger wave function in time-periodic potentials}}
\gdef\shorttitle{Ionization in time-periodic potentials}
\gdef\shortauthors{O. Costin, R. D.  Costin, J. L.  Lebowitz}

\maketitle

{{ \it Dedicated to Elliott Lieb on the occasion of his 70th birthday}}

\bigskip

\begin{abstract}
  We study the transition to the continuum of an initially bound
  quantum particle in $\RR^d$, $d=1,2,3$, subjected, for $t\ge 0$, to
  a time periodic forcing of arbitrary magnitude. The analysis is
  carried out for compactly supported potentials, satisfying certain
  auxiliary conditions. It provides complete analytic information on
  the time Laplace transform of the wave function. From this,
  comprehensive time asymptotic properties (Borel summable
  transseries) follow.
  
  We obtain in particular a criterion for whether the wave function
  gets fully delocalized (complete ionization). This criterion shows
  that complete ionization is generic and provides a convenient test
  for particular cases.  When satisfied it implies absence of discrete
  spectrum and resonances of the associated Floquet operator. As an
  illustration we show that the parametric harmonic perturbation of a
  potential chosen to be any nonzero multiple of the characteristic
  function of a measurable compact set has this property.
  
\bigskip

\z {\bf Keywords:} Ionization, delocalization, resonances, Floquet
theory, Borel summability.

\end{abstract}

\section{Introduction}

We consider the non-relativistic Schr\"odinger equation for the wave
function $\psi(x,t), x\in\RR^d$
\begin{equation}
  \label{eq:eqa} 
  i\, \frac{\partial\psi }{\partial
t}\, =\Big(-\Delta+V(x)+\Omega(x,t)\Big)\, \psi
\end{equation}
where $\Omega(x,t)$ is a time-periodic external potential (not
necessarily small):
\begin{equation}
  \label{eq:eqa2}
 \Omega(x,t)=\Omega(x,t+2\pi/\omega), \ \omega>0
\end{equation}
We take $V$ and $\Omega$ real-valued and so that
\begin{equation}
  \label{as1}
V\in L^\infty(\RR^d),\ \ 
\Omega\in L^\infty(\RR^d\times [0,2\pi/\omega])
\end{equation}
with $\Omega\not\equiv 0$ satisfying
\begin{equation}
  \label{Fourier}
  \Omega(x,t)=\sum_{j\in\ZZ}\Omega_j(x)e^{ij\omega t},\ \ \ \Omega_j(x)=\overline{\Omega_{-j}(x)};\ \ \sup_{j,x}|\Omega_j(x)| j^2<\infty 
\end{equation}
We set, without loss of generality,
\begin{equation}
  \label{Omega0}
  \Omega_0(x)=0
\end{equation}
We are interested in the behavior of solutions $\psi(x,t)$ for large $t$ when
\begin{equation}
  \label{psil2}
   \psi(x,0)=:\psi_0(x)\in L^2(\RR^d), \ \ \ \int_{\RR^d}|\psi_0|^2dx=1
\end{equation}
and $\psi_0$ is sufficiently regular (we assume it of class $C^4$).
Of particular interest is the {\em survival probability} for the
particle in a ball $B$ in $\RR^d$, $\int_{x\in B}|\psi|^2
dx:=\mathcal{P}_B(t)$. If $\mathcal{P}_B(t)$ approaches zero as
$t\rightarrow\infty$ for all $B$, then we say that the particle
escapes to infinity and {\em complete ionization} occurs.

While many results in the paper only require (\ref{as1})
(\ref{Fourier}) plus sufficient algebraic decay of $\Omega$ and $V$
for large $|x|$, some specific results later in the paper,
particularly detailed analytic information, require that $V$ and
$\Omega$ are compactly supported,
\begin{deqarr}
  \text{supp}(V)\cup \text{supp}(\Omega(\cdot,t))\subset {D}
  \label{compactDa} \\\rem{and others require the
    same also for $\psi_0$ in (\ref{psil2})} \nonumber\\
  \text{supp}(\psi_0)\cup \text{supp}(V)\cup
  \text{supp}(\Omega(\cdot,t))\subset {D} \label{compactDb}
 \end{deqarr} 
with 
\begin{equation*}
   {D}\subset\mathbb{R}^d \text{\ compact,} \ 
  \mathbb{R}^d\setminus D \text{\ connected,
    meas}(\partial D)=0
\end{equation*}
The rest of the paper will therefore be
written in the context of this setting.
\subsection{Nature of the results} Under the assumptions (\ref{compactDb})
$\psi(x,t)$ is obtained for large $t$ as a convergent combination of
exponentials and Borel summable power series in $t^{-1/2}$.

If an additional assumption (connected to the absence of discrete
spectrum of the Floquet operator) is satisfied, the long time
expression of $\psi$ contains only decaying terms, cf.
Theorem~\ref{genth12} in \S~\ref{3.8}, i.e. we get complete
ionization.\footnote{While explicit information on long time behavior
  requires $\psi_0$ to be localized, decay for more general $\psi_0\in
  L^2$ is then an immediate consequence of the unitarity of
  Schr\"odinger evolution.}

We find in Proposition~\ref{cdi} a convenient sufficient condition for
complete ionization, and show that it is satisfied by a
nonperturbative example\footnote{The same results hold if a bounded
  time-independent potential, not necessarily constant, with compact
  support disjoint from $D$ is added to $V$, see Remark~\ref{extension}.}
\begin{equation}
  \label{nonp}
  V(x)=V_{{D}}\,\bchi_{{D}}(x);\ \ \Omega(x,t)=2\Omega_{{D}} \,\bchi_{{D}}(x)\sin \omega t
\end{equation}
where $\bchi_{{D}}$ is the characteristic function of $D$ in $d=1,2,3$
and $V_D$ and $\Omega_D$ are arbitrary nonzero constants.

We previously obtained similar results for more general potentials in
$d=1$ and radially symmetric ones in $d=2,3$. See
\cite{JPA,CMP221,CRM,CMP224,genpaper} and \cite{UAB} where there is a
review of our previous work on this problem.

\subsection{Strategy of the approach} The key steps of our approach are outlined in \S\ref{KS}. The method we use  is based on a study of the analytic properties of $\hat{\psi}$, the Laplace transform of $\psi$.
The type and position of the singularities of $\hat{\psi}$, given in
Theorem~\ref{Pn=3} and Lemma~\ref{L12}, provide information about the
time behavior of $\psi$; the former are obtained from an appropriate
equation to which the Fredholm alternative approach applies.

\section{Laplace transform, link with Floquet theory}\label{secLT}

Exisence of a strongly differentiable unitary propagator for
(\ref{eq:eqa}) (see \cite{Reed-Simon} v.2, Theorem X.71) implies that
for $\psi_0\in L^2(\RR^d)$, the Laplace transform
\begin{equation}
  \label{eq:Lap}
  \hat{\psi}(\cdot,p):=\int_0^{\infty}\psi(\cdot,t)e^{-pt}dt
\end{equation}
exists for $\Re(p)>0$. It satisfies the equation
\begin{equation}
  \label{eq:S-lt}
(-\Delta+V(x)-ip)\hat{\psi}(x,p)=-i\psi_0-\sum_{j\in\ZZ} \Omega_j(x)\hat{\psi}(x,p-ij\omega)
\end{equation}
and the map $p\to \psi(\cdot,p)$ is $L^2$ valued analytic in the right
half plane
\begin{equation}
  \label{H}
  p\in\mathbb{H}=\{z:\Re(z)>0\}
\end{equation}
Clearly, equation (\ref{eq:S-lt}) couples $\hat{\psi}(x,p_1)$ with
$\hat{\psi}(x,p_2)$ iff $(p_1-p_2)\in i\omega\ZZ$. Setting
\begin{equation}
  \label{defnp}
  p=i(\sigma+n\omega)\text{ with }
\Re\,\sigma\in[0,\omega)
\end{equation}
(sometimes it will be technically helpful to relax this restriction on
$\sigma$) we define $y_n^{[1]}(x;\sigma)=
\hat{\psi}(x,i(\sigma+n\omega))$. Eq (\ref{eq:S-lt}) now becomes a
differential-difference system
\begin{equation}
  \label{eq:S-lt2}
(-\Delta+V+\sigma+n\omega)y_n^{[1]}=-i\psi_0-\sum_{j\in\ZZ}\Omega_j(x)\left(S^{-j}y^{[1]}\right)_n
\end{equation}
where the shift operator $S$ is given by
\begin{equation}
  \label{Sdef}
  (Sy)_n=y_{n+1}
\end{equation}
\subsection{Connection with Floquet theory}\label{cF} The solution of (\ref{eq:eqa})
with time periodic $\Omega$ is of course the subject of Floquet theory
(see \cite{[30]} and \cite{YajPriv}-\cite{Reed-Simon}) and therefore
our analysis connects to it in a number of ways. Let $K$ be the
quasi-energy operator in Floquet theory
\begin{equation}
  \label{eq:quasien}
 (Ku)(x,\theta)= \left(-i\frac{\partial}{\partial \theta} -\Delta+V(x)+\Omega(x,\theta)\right)u(x,\theta); \ x\in \RR^d,\,\theta\in S^1_{{2\pi}/{\omega}}
\end{equation}
Then, letting
\begin{equation}
  \label{ftu}
  u(x,\theta;\sigma)=\sum_{n\in\ZZ}y_n^{[1]}(x;\sigma)e^{in\omega\theta}
\end{equation}
be the solution of the eigenvalue equation
\begin{equation}
  \label{evK}
  Ku=-\sigma u
\end{equation}
we get an equation for the $y_n^{[1]}$ which is identical to the
homogeneous part of equation (\ref{eq:S-lt2})\footnote{The functional
  spaces are different.  Proposition~\ref{Fl3} clarifies this
  question.}. Solutions of (\ref{evK}) with $u\in L^2(\RR^d\times
S^1_{{2\pi}/{\omega}})$ correspond to eigenfunctions of $K$.
\begin{Remark}\label{R01}
  If $u$ is an eigenfunction of $K$ corresponding to the eigenvalue
  $-\sigma$, then $ue^{-ij\omega\theta}$ is an eigenfunction with
  eigenvalue $-\sigma+j\omega$.  For this reason it is enough to
  restrict $\sigma$ to the strip given in (\ref{defnp}).

\end{Remark}
Complete ionization clearly requires the absence of a discrete
spectrum of (\ref{evK}) (Otherwise, if $u(x,\theta)$ is an
eigenfunction of $K$, then $e^{i\sigma t}u(x,t)$ would be a space
localized solution of the Schr\"odinger equation.)  In a recent work
\cite{YajPriv}, Galtbayar, Jensen and Yajima proved that the opposite
is also true.  They obtained asymptotic series in $t^{-1/2}$ for the
projection of the wave function $\psi(x,t)$ on the space orthogonal to
the discrete spectrum of $K$.

Our approach via Laplace transform is different from that of
\cite{YajPriv}. For the corresponding time evolution our results are
stronger than those obtained in \cite{YajPriv} but apply to the more
restrictive classes of $V$ and $\Omega$ satisfying (\ref{compactDb})
in $d=1,2,3$. We show that the time behavior of $\psi(x,t)$ is given
by a Borel summable transseries containing both power law decay and
exponential terms. For potentials satisfying our condition
(\ref{nonp}) we show that $K$ has no discrete spectrum or resonances
and all the exponentials are decaying.

More details on the connection between our work, Floquet theory and
\cite{YajPriv} are given in \S\ref{F22}.

\section{Integral equation, compactness and analyticity}
\subsection{Laplace space equation} 

To simplify contour deformation, we first improve the decay of
$\hat{\psi}$ for large $p$, by pulling the first two terms in the
asymptotic behavior for large $p$ from $\hat{\psi}$.  Let $\delta_{ij}=1$ if $i=j$, and $0$ otherwise,   $\delta_{ij}^c=1-\delta_{ij}$ and define the
operator $\mathfrak{N}$ by
\begin{equation}
  \label{deffrakN}
 (\mathfrak{N}f)_n(x)=(-\Delta+V)\frac{\delta^c_{n0} f_n(x)}{\sigma+n\omega}-\delta_{n0}f_{n,x}+\sum_{k\ne n}\Omega_{n-k}(x)\frac{\delta^c_{k0} f_k(x)}{\sigma+k\omega}
\end{equation}
\subsection {Green function representation}
To pass to an integral form of the system of equations (\ref{eq:finy})
we apply to them the Green function of $(-\Delta+\sigma+n\omega)$,
given by
\begin{equation}
  \label{GreenF}
    \Big(\mathfrak{g}_nf\Big)(x)=\int G(\kappa_n(x-x')) f(x')dx'
\end{equation}
with 
\begin{equation}
  \label{defkappa}
  \kappa_n=\sqrt{-ip}=\sqrt{\sigma+n\omega}\  \text{(when
$p\in\mathbb{H}$, $\kappa_n$ is in the fourth quadrant)} 
\end{equation}
and
\begin{equation}
  \label{formulares}
  G(\kappa_nx)=\left\{
\begin{array}{ccccc}\displaystyle \frac{1}{2}\kappa_n^{-1}e^{-\kappa_n|x|}\ \ & d=1\\ & \\ \displaystyle 
\frac{1}{2\pi}K_0(\kappa_n|x|)\ \  & d=2\\ &\\ \displaystyle 
\frac{1}{4\pi}|x|^{-1}e^{-\kappa_n|x|}\ \ & d=3 \end{array}\right.
\end{equation}
(see \cite{Reed-Simon}) where $K_0$ is the modified Bessel function of
second kind,
\begin{equation}
  \label{Bessel0}
  K_{0}(x)=\int_0^{\infty}e^{-x\cosh t}dt=e^{-x}\int_0^{\infty}\frac{e^{-xs}}{\sqrt{s(s+2)}}ds
\end{equation}
Note that, in the setting (\ref{compactDa}), for $f$ supported in
${D}$ we have
\begin{equation}
  \label{GreenF2}
    \Big(\mathfrak{g}_nf\Big)(x)=\int_{{D}} G(\kappa_n(x-x')) f(x')dx'
\end{equation}
Eq. (\ref{deffrakN}) in integral form becomes
\begin{equation}
  \label{eq:E}
  y_n^{[1]}=-i\mathfrak{g}_n\psi_0+(\mathfrak{C}y^{[1]})_n
\end{equation}
where 
\begin{equation}
  \label{eq:E1}
  y^{[1]}=(y_n^{[1]})_{n\in\ZZ}\ \ \text{and}\ \ (\mathfrak{C}y^{[1]})_n=-\mathfrak{g}_n\left[Vy_n^{[1]}+\sum_{j\in\ZZ}\Omega_j(x)\left(S^{-j}y^{[1]}\right)_n\right]
\end{equation}
To ensure better decay with respect to $n$ we further substitute in (\ref{eq:E1}) 
\begin{equation}
  \label{eq:E2}
  y_n^{[1]}=-i\mathfrak{g}_n\psi_0-i\psi_{1,n}+y_n
\end{equation}
where $(\psi_{1,n})_{n\in\ZZ}=:\psi_1$ and
\begin{equation}
  \label{eq:E4}
  \psi_1=\mathfrak{C}\left[\left(\mathfrak{g}_n\psi_0\right)_{n\in\ZZ}\right]
\end{equation}
Then $y$ satisfies
\begin{equation}
  \label{eq:fint2}
 y=w+\mathfrak{C}(\sigma)y
\end{equation}
(We write $\mathfrak{C}$ for $\mathfrak{C}(\sigma)$ when the
dependence on $\sigma$ need not be stressed.) In differential form
(\ref{eq:fint2}) reads
\begin{equation}
  \label{eq:finy}
  (-\Delta+\sigma+n\omega)y_n=i\psi_{2,n}-Vy_n-
\sum_{j\in\ZZ}\Omega_j(x)\left(S^{-j}y\right)_n
\end{equation}
\subsection{The Hilbert space} To analyze the properties of (\ref{eq:fint2}) we use the Hilbert space
\begin{equation}
  \label{defH}
  \mathcal{H}=l^2_{\gamma}(L^2(B))
\end{equation}
where $B$ is an arbitrary ball (containing ${D}$) defined as the space
of sequences $\{y_n\}_{n\in\ZZ},\,y_n\in L^2(B)$ with
$$\|y\|^2_{\mathcal{H}}=\sum_{n\in\ZZ}|n|^\gamma\|y_n\|^2_{L^2(B)}<\infty$$
and adequate $\gamma$; we take for definiteness $\gamma=3/2$; larger
$\gamma$ can be taken if one assumes more differentiability than
(\ref{Fourier}) implies.  (Note that $\mathcal{H}$ is different from
the Hilbert space $L^2(L^2(\RR^d))$ used in Floquet theory. See also
\S\ref{F22}.)
\subsection{Strategy of the approach, continued}\label{KS}
As mentioned, unitarity of the evolution shows that
$\hat{\psi}(\cdot,p)\in L^2(\RR^d)$ if $\Re(p)>0$.  In the integral
form (\ref{eq:fint2}), whose solutions are in $\mathcal{H}$ when
$\Re(p)>0$, the operator is, under our assumptions, compact.  The
solution of this equation is shown to be unique in $\mathcal{H}$ for
large enough $\Re(p)$ by the contractivity of the integral operator.
Uniqueness and analyticity of the solution for $p$ in the right half
plane $\mathbb{H}$ follow by an application of the analytic version of
the Fredholm alternative \cite{Reed-Simon}.  We then show that the
solution and thus $\hat{\psi}$ are analytic with respect to a
uniformizing variable, in appropriately chosen domains containing
parts of the imaginary axis.  The contour of the inverse Laplace
transform, $\tau\to c+i\tau;\tau\in\RR, c>0$, can then be deformed to
$i\RR$ (the boundary of $\mathbb{H}$) where $\hat{\psi}$ is analytic
except for a discrete set of square root branch points.  The large
time behavior of $\psi$ follows.
\subsection{Compactness}
In $d=1$, we further transform the equation, see (\ref{d1,2}) in
Appendix~\ref{pd=1}, to improve the regularity of the operator at
$n=0$ and $\sigma=0$.
\begin{Lemma}\label{L4} Under the assumptions (\ref{compactDa}), $w\in\mathcal{H}$ and 
  $\mathfrak{C}$ is a compact operator on $\mathcal{H}$.
\end{Lemma}
To show that $w\in\mathcal{H}$ we use the fact that the operators $\mathfrak{g}_n$ satisfy (see Appendix A of \cite{Agmon}, and also \S\ref{compactness}).
\begin{equation}
  \label{eq:Agmon}
 \sup_{n\in\ZZ}(1+|n|)^{1/2}\|\mathfrak{g}_n\|_{L^2(\mathcal{D})}<\infty
\end{equation}
Then
\begin{equation}
  \label{eq:E6}
\sup_{n\in\ZZ} (1+|n|)^{1/2} \|(\mathfrak{g}_n\psi_0)_{n}\|<\infty\ \ \text{implying} \ \ 
\sup_{n\in\ZZ}(1+|n|)\|\psi_{1,n}\|_{L^2(\mathcal{D})}<\infty
\end{equation}
In view of (\ref{Fourier}) we also have
\begin{equation}
  \label{eq:E8}
  \sup_{n\in\ZZ}(1+|n|)\left\|V\psi_{1,n}+\sum_{j\in\ZZ}\Omega_j(x)\left(S^{-j}y\right)_n\right\|_{L^2(\mathcal{D})}<\infty
\end{equation}
implying
\begin{equation}
  \label{eq:E9}
  \sup_{n\in\ZZ}(1+|n|^{3/2})\|w_n\|_{L^2(\mathcal{D})}<\infty
\end{equation}

It is not difficult to check that
$\mathfrak{g}_n$ is compact on $L^2(B)$ for each $n$; it is
more delicate to show compactness of $\mathfrak{C}$; both properties
are proved in \S\ref{S6}.
\subsection{Uniqueness}
\begin{Lemma}\label{L5}
  For large enough $-\Im\sigma$, eq. (\ref{eq:fint2}) has a unique
  solution in $\mathcal{H}$.
\end{Lemma}
\z The proof is given in Appendix~\ref{Unic}.

\subsection{Analytic structure of  $\hat\psi$}
\begin{Remark}\label{R4}
  It is convenient to introduce the uniformizing variable
  $\sigma=u^2$; with the natural branch of the square root, $u$ is in
  the fourth quadrant when $\sigma$ is in the lower half plane.  In
  this variable, we write $\kappa_0=u$ and
  $\kappa_n=\sqrt{n\omega+u^2}$ for $n\ne 0$.
\end{Remark}
\begin{Proposition}\label{L6}
  In the setting (\ref{compactDa}) the operator
  $\mathfrak{C}(\sigma)$ is analytic in $u$ in the region
  $S_{\omega}=\{u:|\Re u^2|<\omega\}$ hence in $\sqrt{\sigma}$
  in the strip (see Remark~\ref{R01})
\begin{equation}
  \label{strip}
\Big\{\sigma:\Re(\sigma)\in(-\omega,\omega)\Big\}
\end{equation}
Additionally, $\mathfrak{C}(\sigma)$ is analytic in $\sigma$ at any
$\sigma\ne 0$.
\end{Proposition}
\z {\em Proof.}  In terms of $u$ we define $\kappa_0=u$,
$\kappa_n=\sqrt{u^2+n\omega}$ and then $\kappa_n$ is analytic in $u$
in the simply connected region $S_\omega$ for any $n\in\ZZ$. Since in
the setting (\ref{compactDa}) the integral (\ref{GreenF2}) is over
a compact set, $D$, $\mathfrak{g}_n$ is also analytic for any
$n\in\ZZ$. Analyticity of $\mathfrak{C}$ follows from the fact that
\begin{equation}
  \label{limC}
  \mathfrak{C}=\lim_{N\to\infty} \mathfrak{C}_N\
\end{equation}
with
\begin{equation}
  \label{decC}
   \Big(\mathfrak{C}_N\,\,y\Big)_n :=\left\{\begin{array}{cc}\ \ \big(\mathfrak{C}y\big)_n; \ \ &|n|\le N\\ 0  &{\text otherwise}\end{array} \right.
\end{equation}
and convergence is uniform in $u$ on compact subsets of $S_{\omega}$.
This is shown in  Lemma~\ref{L27}.  $\Box$

\begin{Proposition}\label{P4} There exists a unique solution $y$ to (\ref{eq:fint2})
  and it has the same analyticity properties as $\mathfrak{C}$ if
 \begin{equation}
  \label{hmg}
\text{$ ${\bf(A)} \text{ For $\Im \sigma \le 0$}\ \ \ 
    ($  v=\mathfrak{C}v$,  $v\in\mathcal{H}$)}\ \Rightarrow v=0 
\end{equation}
\end{Proposition}
\z {\em Proof}. This is nothing more than the analytic Fredholm
alternative (see e.g. \cite{Reed-Simon} Vol 1, Theorem VI.14, pp. 201).
$\Box$

This formulation is convenient in determining the analytic properties
of $y$ with respect to $\sigma$, instrumental for the Borel
summability results stated in Proposition~\ref{genth1}.

\begin{Theorem}\label{Pn=3}
  If {\bf (A)} and (\ref{compactDb}) hold, then: 
  
  (i) the solution $y$ of (\ref{eq:fint2}) is meromorphic in $u$ in
  the disk $\{u:|u|<\sqrt{\omega}\}$, see Remark~\ref{R4}, analytic at
  $u=0$ and in the fourth quadrant of $S_{\sqrt{\omega}}$.
  Furthermore, $y$ is analytic in $\sigma$ at any $\sigma_0\ne 0$.
 
  (ii) $\hat{\psi}$ is analytic in $p$ in a cut neighborhood of
  $i\RR$, $\{p:\Re(p)>-\epsilon\}$ with cuts toward $-\infty$ at $i
  \,n\,\omega,\ n\in\ZZ$.  Furthermore, in a neighborhood of
  $i\,n\,\omega$, see (\ref{defnp}), we have
$\hat{\psi}(p)=A_n(p)+B_n(p)\sqrt{\sigma}$ where $A_n$ and $B_n$ are
analytic at $i\,n\,\omega$ and, for some $\epsilon<\sqrt{\omega}$ and
$|u|\le \epsilon$ we have
\begin{equation}
  \label{normAB}
  \sup_{n\in\ZZ,|u|<\epsilon}n^{\gamma}\Big(\|A_n(p)\|_{L^2(B)}+\| B_n(p)\|_{L^2(B)}\Big)<\infty
\end{equation}
\end{Theorem}

\z {\em Proof.}  (i) follows from Propositions~\ref{L6} and
~\ref{P4}, and from the link between $\hat{\psi}$ and $y_n$.

(ii) The functions $A_n$ and $B_n$, are simply the even and odd part
respectively of the analytic function $y_n(x;u)$.  The estimate
follows from the fact that  $y(x;u)\pm
y(x;-u)\in\mathcal{H}$ is analytic in $u$ for $|u|<\epsilon$.

In Proposition~\ref{genth1} and  Theorem~\ref{genth12} below we use the following result.
\begin{Lemma}\label{L12}
  In the setting (\ref{compactDa}), if $\hat{\psi}$ has a pole at
  $\sigma=\sigma_0\in i\RR$, then, in the variable $\sigma-\sigma_0$
  if $\sigma_0\ne 0$ or $u$ if $\sigma_0=0$, the pole is simple.
\end{Lemma}
This is shown in Appendix \ref{simpoles}. $\Box$
\begin{Proposition}\label{Pex} Condition
  {\bf (A)} is satisfied for the potentials (\ref{nonp}).
\end{Proposition}
\z {\em Proof.}  This is  established in \S \ref{Examples}. $\Box$
\subsection{Asymptotic expansion of $\psi$ and  Borel summability}\label{3.8}
\begin{Proposition}\label{genth1}
  In the setting (\ref{compactDb}) there exist $N\in\NN$,
  $\{\Gamma_k\}_{k\le N}$, and $\{F_{\omega;k}(t,x)\}_{k\le N}$,
  $2\pi/\omega$-periodic functions of $t$, such that, for $t>0$,

\begin{equation}\label{transP3}\psi(t,x)=\sum_{j\in\ZZ}e^{ij\omega
t}h_j(t,x)+\sum_{k=1}^N P_k(t)e^{-\Gamma_k t}F_{\omega;k}(t,x)\end{equation}
with $\Re\Gamma_k\ge 0$ for all $k\le N$,  $P_k(t)$ are polynomials
in $t$, reducing to constants if
$\Re\Gamma_k=0$, and the 
$h_j(t,x)$ have Borel summable power series in $t^{-1/2}$
\begin{equation}\label{e113}h_j(t,x)=\mathcal{LB}\sum_{k\ge k_0}h_{kj}(x)t^{-k/2}
\end{equation}
with $k_0\ge 1$.
\end{Proposition}

\begin{Remark}[Borel summation] \label{F2}  If $\tilde{f}$ is a formal power series, say in inverse 
  powers of $t$, then $\tilde{F}=\mathcal{B}\tilde{f}$ is also a
  formal power series, defined as the term-wise inverse Laplace
  transform in $t$ of $\tilde{f}$. If (1) $\tilde{F}$ is convergent
  (2) its sum $F$ can be analytically continued along $\RR^+$ and (3)
  $\exists \nu$ s.t. $F(p)\in L^1(\RR^+, e^{-\nu p}dp)$, then the
  Laplace transform $\mathcal{L}F$ is by definition the Borel sum of
  $\tilde{f}$ denoted by $\mathcal{LB}\tilde{f}$. In our context we
  have, more precisely,
    $$h_j(t,x)=\int_0^\infty\, F_j(\sqrt{p},x)\, e^{-pt}\, dp\, \sim
    \, \sum_{k}\, h_{kj}(x)\, {t^{-k/2}},\ t\to+\infty$$
  \z where the functions $F_j(s,x)$ are analytic in $s$ in a
    neighborhood of $\RR^+$ and for any $b\in\RR$ there exist
   a  constant $C$ such that for all $j$ and $p\in\RR^+$,
   $$\sup_{p\ge 0;|x|<b}|F_j(\sqrt{p},x)e^{-C|p|}| \le f_j$$
   where the
   $f_j$ decay in $j$ faster than $j^{-2}$ under the assumption
   (\ref{Fourier}) and factorially if $\Omega$ is a trigonometric
   polynomial. Thus the function series in (\ref{transP3}) converges
   (rapidly in the latter case).
    \end{Remark}

\z     The role of condition {\bf (A)} is described in the following
    result.

\begin{Theorem}\label{genth12}
  (i) If {\bf (A)} holds, then on the right side of (\ref{e113}) and
  (\ref{transP3}) we have
\begin{equation}
  \label{Aholds}
  k_0\ge 3\ \text{ and }\ \Re\Gamma_k> 0\ \text{ for all }\ k.
\end{equation}
In particular we have complete ionization of the system. (See also
Proposition~\ref{Pex}, as well as Proposition~\ref{cdi} and
Remark~\ref{Overdet} below.)

(ii) If {\bf (A)} is not satisfied, then some $\Re\Gamma_k$ may
vanish; the part of $\psi$ corresponding to these $\Gamma_k$ remains a
spatially localized quasiperiodic function of $t$. Exceptionally,
$k_0=1$ if {\bf (A)} does not hold (see also \cite{YajPriv} and
Proposition~\ref{Fl3}).
\end{Theorem}

\z The proofs of Proposition~\ref{genth1} and Theorem~\ref{genth12}
are sketched in Appendix~\ref{Sketch}. In one dimension a similar
result is stated in \cite{UAB}.
 \section{Ionization condition for compactly supported potentials} 
For the setting (\ref{compactDa}) we derive a technically convenient
condition implying {\bf (A)}.  

Assume $0\ne v\in\mathcal{H}$ and $ v=\mathfrak{C}v$. Then there
exists a nontrivial solution in $\mathcal{H}$ to the system
\begin{equation}
  \label{hmdiff}
  (-\Delta+\sigma+n\omega)y_n=-Vy_n-\sum_{j\in\ZZ}\Omega_j(x)y_{n-j}
\end{equation}
We multiply (\ref{hmdiff}) by $\overline{y}_n$, integrate over a ball
$B$ containing ${D}$, sum over $n$ (which is legitimate since
$y\in\mathcal{H}$) and take the imaginary part of the resulting
expression. Noting that
\begin{multline}
  \label{conjsym}
  \overline{\sum_{j,n\in\ZZ}\Omega_j(x)y_{n-j}\overline{y_n}}=\sum_{j,n\in\ZZ}
\Omega_{-j} \overline{y}_{n-j}y_{n}=\sum_{j,n\in\ZZ}
\Omega_{j} \overline{y}_{n+j}y_{n}\\=\sum_{j,m\in\ZZ}\Omega_j(x)\overline{y_m}y_{m-j}\end{multline}
so the sum (\ref{conjsym}) is real, we get
\begin{multline}
  \label{eq:nonreal}
  0=\Im\left(-\sigma\sum_{n\in\ZZ}\|y_n\|^2+\int_{B}\sum_{n\in\ZZ} dx
    \overline{y}_n\Delta y_n\right)\\=
  -\Im\sigma\sum_{n\in\ZZ}\|y_n\|^2+\frac{1}{2i} \int_{\partial
    B}\left(\sum_{n\in\ZZ}\overline{y}_n\nabla y_n-y_n\nabla
    \overline{y}_n\right)\cdot \mathbf{n}\,dS\end{multline} 

We take
$d=3$ (the analysis is simpler in one or two dimensions). It is
convenient to decompose $y_n$ using spherical harmonics; we write
\begin{equation}
  \label{sph}
  y_n=\sum_{l\ge 0, |m|\le l}R_{n,l,m}(r)Y_l^m(\theta,\phi).
\end{equation}
The last integral in (\ref{eq:nonreal}), including the prefactor, then
equals
\begin{multline}
  \label{sph2}
  -8\pi\,i\, r_B^2\sum_{n\in\ZZ}\sum_{m,l}\Big[\overline{R}_{n,m,l}R'_{n,m,l}-
  \overline{R'}_{n,m,l}R_{n,m,l}\Big]\\=-8\pi\,i\,  r_B^2\sum_{n\in\ZZ}\sum_{m,l}W[\overline{R}_{n,m,l},R_{n,m,l}]
\end{multline}
where $r_B$ is the radius of $B$ and $W[f,g]$ is the Wronskian of $f$
and $g$.  On the other hand, since $V$ and $\Omega$ are compactly
supported, we have outside of $B$
\begin{equation}
  \label{eq:outside}
  \Delta y_n-(\sigma+n\omega)y_n=0
\end{equation}
and then by (\ref{sph}), $R_{n,l,m}$ satisfy for $r>r_B$ the equation
\begin{equation}
  \label{Rnlm}
  R''+\frac{2}{r}R'-\frac{l(l+1)}{r^2}R=(\sigma+n\omega)R
\end{equation}
where we have suppressed the subscripts. Let $g_{n,l,m}=rR_{n,l,m}$.
Then for the $g_{n,l,m}$ we get
\begin{equation}
  \label{gnlm}
  g''-\left[\frac{l(l+1)}{r^2}+(\sigma+n\omega)\right]g=0
\end{equation}
thus
\begin{equation}
  \label{r-g}
  \overline{R}R'=\frac{\overline{g}g'}{r^2}-\frac{|g|^2}{r^3}
\end{equation}
and 
\begin{equation}
  \label{r-g1}
  r^2W[\overline{R},R]=W[\overline{g},g]=:W_n. 
\end{equation}
Multiplying (\ref{gnlm}) by $\overline{g}$, the conjugate of
(\ref{gnlm}) by $g$ and subtracting, we get for $r>r_B$,
\begin{equation}
  \label{difg}
  W_n'=(\sigma-\overline{\sigma})|g|^2=2i|g|^2 \Im\,\sigma 
\end{equation}

\begin{Remark}\label{condifty}
  Simple estimates using equation (\ref{hmg}), the definition
  (\ref{GreenF}) and (\ref{defkappa}) imply that, for some $c_n$,
\begin{equation}
  \label{cinf}
  y_n(x)=\frac{e^{-\kappa_n|x|}}{|x|}\Big(c_n(\theta,\phi)+O(|x|^{-1})\Big)\ \text{as}\ |x|\to\infty
\end{equation}
\end{Remark}

\smallskip

\z Let us consider two cases of (\ref{hmdiff}).

\z Case (i): $\Im\,\sigma\,<0$. By Remark \ref{condifty} we have
\begin{equation}
  \label{ginfty}
  g\sim C e^{-\kappa_n r}(1+o(1))\ \ \text{as}\ \ r\to\infty
\end{equation}
There is a one-parameter family of solutions of (\ref{gnlm})
satisfying (\ref{ginfty}) and the asymptotic expansion can be
differentiated \cite{Wasow}. We assume, to get a contradiction, that
there exist $n$ for which $g_n\ne 0$.  For these $n$ we have, using
(\ref{ginfty}), differentiability of this asymptotic expansion and
(\ref{defkappa}) that
\begin{equation}
  \label{t2}
  \frac{1}{2i}\lim_{r\to\infty}|g_n|^{-2} W_n=-\Im\kappa_n\, >0.
\end{equation}
It follows from (\ref{difg}) and (\ref{t2}) that $\frac{1}{2i}W_n$ is
strictly positive for all $r>r_B$ and all $n$ for which $g_n\ne
0$.  This implies that the last term in (\ref{eq:nonreal}) is a sum of
positive terms which shows that (\ref{eq:nonreal}) cannot be satisfied.

Case (ii): $\Im\,\sigma\,=0$. For $n>0$ there exists only one solution
$g$ of (\ref{gnlm}) which decays at infinity (cf. Remark
\ref{condifty} and the discussion in Case (i)), and since (\ref{gnlm})
has real coefficients this $g$ must be a (constant multiple of a) real
function as well; therefore we have $W_n=0$ for $n\ge 0$.

For $n<0$, we use Remark \ref{condifty} (and differentiability of the
asymptotic expansion as in Case (i)) to calculate the Wronskian $W_n$
of $g,\overline{g}$ in the limit $r\to \infty$: $W_n=|c_n|^2
(1+o(1))$.  Since for $\Im\,\sigma=0$, $W_n$ is constant, cf.
(\ref{difg}), it follows that $W_n$ is exactly equal to $|c_n|^2$.
Thus, using (\ref{eq:nonreal}) and (\ref{sph2}) we have
\begin{equation}
  \label{IonizCond}
  y_n(x)=0\ \ \text{for all}\ \ n<0 \ \text{and}\ |x|>r_B
\end{equation}
\begin{Proposition}\label{cdi}
  In the setting (\ref{compactDa}), if {\bf (A)} fails, then we have
  \begin{equation}
    \label{I2}
   y_n(x)=0\ \ \text{for all}\ \ n<0 \ \text{and}\ x\notin {D}
  \end{equation}
\end{Proposition}
Outside $D$ we have $\mathfrak{O}y_n=0$, where $\mathfrak{O}$ is the
elliptic operator $-\Delta+\sigma+n\omega$. The proof follows
immediately from (\ref{IonizCond}), by standard unique continuation
results \cite{Hormander}, \cite{Miranda}, \cite{Treves} ( in fact,
$\mathfrak{O}$ is analytic hypo-elliptic).
 \begin{Remark}\label{Overdet}
   Proposition~\ref{cdi} points toward generic ionization under time
   periodic forcing. Indeed, we see from (\ref{I2}) that equations
   (\ref{hmdiff}) are formally overdetermined when $n<0$ ($y_n$ is in
   the domain of $\Delta$ so that, in (\ref{hmdiff}), the function and
   ``one derivative'' are given on the boundary) and are expected,
   generically, not to have nontrivial solutions even if $y_{n}$ 
   had to satisfy (\ref{eq:finy}) for $n<0$ only.
 
   The latter reduced problem is relatively easier to study and we
   used it to show that {\bf (A)} holds in a number of settings,
   including the potential (\ref{nonp}), see \cite{UAB}.
\end{Remark}
\begin{Remark}
  There do in fact exist nongeneric potentials (though not in the
  class (\ref{as1})) for which ionization fails
  \cite{CMP221,[6a],RCL}.
\end{Remark}

\section{Connection with Floquet theory, continued}\label{F22}
\begin{Proposition}\label{P17}
  If $u$ is an eigenfunction or resonance\footnote{See \cite{YajPriv}
    and Proposition~\ref{Fl3} (ii).} of the operator $K$ defined in
  (\ref{eq:quasien}) such that $u\in L^2(\RR^3\times S^1_{{2\pi}/{\omega}})$, then
  $u=\mathfrak{C}u$ in $\mathcal{H}$ and so {\bf (A)} fails.
  \end{Proposition}
  The proof is an immediate consequence of Proposition~\ref{P161} in
  Appendix~\ref{PP21}.  Conversely, we have the following result.
\begin{Proposition}\label{Fl3} We assume  the setting (\ref{compactDa}). 
  
  (i) If $v=\mathfrak{C}v$ for some $\sigma_0\in(0,\omega)$ and
  $v\in\mathcal{H}$, then $v\in l^2(L^2(\RR^d))$ thus it is an
  eigenfunction of $K$.
  
  (ii) If $v=\mathfrak{C}v$ for $\sigma_0=0$ and $v\in\mathcal{H}$,
  then, for $d=3$, $v$ is of the form
  $v=C|x|^{-1}\delta_{n0}+\tilde{v}(x)$ where $\tilde{v}\in
  l^2(L^2(\RR^3))$ (resonance of $K$).
  
\end{Proposition}

\z {\em Proof of Proposition~\ref{Fl3}.}  For the same reasons as before, we focus on $d=3$.

(i) We see from (\ref{GreenF2}), (\ref{formulares}) and (\ref{I2})
that for all $n$ we have $y_n\in L^2(\RR^3)$.  Furthermore, a
straightforward calculation shows that $\|y_n\|_{L^2(\RR^3)}\le
C\|y_n\|_D $ where $C$ is independent of $n$.  Proposition~\ref{P161}
in Appendix~\ref{PP21} gives the necessary estimates in $n$ to
complete the proof in this case.

(ii) For $n\ne 0$ we have, for the same reasons as in (i), $y_n\in
L^2(\RR^3)$.  But now, at $n=0$, since $\sigma_0=0$ the Green function
(\ref{formulares}) does not have enough decay to ensure $y_0\in
L^2(\RR^3)$.  We have instead, for $x'\in D$ and $|x|\to\infty$,
$G_0(x-x')=\frac{1}{4\pi}|x|^{-1}+O(|x|^{-2})$.  The statement now
follows from (\ref{GreenF2}) and (\ref{eq:fint2}).  $\Box$

\section{Example (\ref{nonp})}\label{Examples} To show that it can be effectively checked whether (\ref{I2}) can be nontrivially satisfied, we consider the example (\ref{nonp}). 
It is convenient to Fourier transform the system (\ref{hmdiff}) in
$x$. In view of (\ref{I2}), for $n<0$, $y_n=0$ outside ${D}$.  We then
have, for $n<0$,
\begin{equation}
  \label{eq:FT1}
  \check{y}_n:=\int_{\RR^3} y_n e^{-ik\cdot x}dx=\int_{{D}} y_n e^{-ik\cdot x}dx
\end{equation}
and
\begin{equation}
  \label{eq:FT}
-k^2\check{y}_n=-k^2\int_{\RR^3} y_n e^{-ik\cdot x}dx=  \int_{\RR^3}\Delta y_n e^{-ik\cdot x}dx=\int_{{D}}\Delta y_n e^{-ik\cdot x}dx
\end{equation}
For the setting (\ref{nonp}) and $n<-1$, (\ref{hmdiff}) reads
\begin{equation}
  \label{F1}
  (k^2+\sigma+n\omega)\check{y}_n=-V_{{D}}\check{y}_n+i\Omega_{{D}}\left( \check{y}_{n+1}- \check{y}_{n-1}\right)
\end{equation}
\begin{Remark}\label{R12}
  For $n\le -1$, the functions $\check{y}_n$ are entire of exponential
  order one; more precisely, if $B$ is a ball containing ${D}$
  we have
  \begin{equation}
    \label{expord1}
    |\check{y}_n(k)|\le  \sqrt{{\rm Vol}({D})} \,\,e^{|k|r_B}\,\,\|y_n\|_{L^2(D)}
  \end{equation}
\end{Remark}
\z {\em Proof.} This follows immediately from the definition of
$\check{y}$. (See also \cite{Zemanian} for a comprehensive
characterization of the Fourier transform of a compactly supported
distribution.)
\begin{Proposition}\label{entire}
  The generating function
\begin{equation}
  \label{eq:defY}
  Y(k,z)=\sum_{m\ge 0}\check{y}_{-m-2}(k)z^m
\end{equation}
is entire in $k$ and analytic in $z$ for $|z|<1$.
\end{Proposition}
\z {\em Proof}.  Since $y\in\mathcal{H}$ we have
\begin{equation}
  \label{bound}
  \|y_n\|_{L^2(D)}\le {\rm const}\,\, |n|^{-3/2}
\end{equation} Using Remark~\ref{R12} the conclusion follows.

A straightforward calculation shows that $Y$ satisfies the equation
\begin{equation}
  \label{eqY}
  M Y-z \frac{\partial Y}{\partial z}-i\beta \left(z
-\frac{1}{z} \right)Y=
i\beta\check{y}_{-1} +i\beta\frac{\check{y}_{-2}}{z}
\end{equation}
where 
\begin{equation}
  \label{defM}
M=\omega^{-1}(k^2+\sigma-2\omega+V_{{D}})
\end{equation}
and $\beta=\Omega_{{D}}/\omega$.
The solution of (\ref{eqY}) is
\begin{equation}
  \label{sol1}
  Y=z^Me^{-i\beta(z+z^{-1})}\left[C(k)-i\beta\int_0^z e^{i\beta(s+s^{-1})}
\left(\frac{\check{y}_{-1}}{s^{M+1}}
+\frac{\check{y}_{-2}}{s^{M+2}}\right)ds\right]
\end{equation}
where the integral follows a path in which $0$ is approached along the
negative imaginary line.
\begin{Remark}
  Proposition~\ref{entire} implies $C(k)\equiv 0$.
\end{Remark}
\z {\em Proof}. It is easy to check that otherwise the limit of
$Y(k,z)$ as $z\to 0$ along $i\RR^{-}$ would not exist. $\Box$

Thus 
\begin{equation}
  \label{sol2}
  Y(k,z)=-i\beta z^Me^{-i\beta(z+z^{-1})}\int_0^z  e^{i\beta(s+s^{-1})}
\left(\frac{\check{y}_{-1}(k)}{s^{M+1}}
+\frac{\check{y}_{-2}(k)}{s^{M+2}}\right)ds
\end{equation}
We now use the nontrivial monodromy of $Y$ on the Riemann surface of
$\log z$, following from the integral representation (\ref{sol2}).
Analytic continuation around the origin gives
\begin{equation}
  \label{monodr}
  i\beta^{-1}e^{i\beta(z+z^{-1})} (Y(\cdot,ze^{2\pi i})-Y(\cdot,z))=\check{y}_{-1}\oint_{\mathcal{C}} \frac{e^{i\beta(s+s^{-1})}}{s^{M+1}}ds+\check{y}_{-2}\oint_{\mathcal{C}} \frac{e^{i\beta(s+s^{-1})}}{s^{M+2}}ds
\end{equation}
where $\mathcal{C}$ is the curve shown in Fig. 1.
\begin{figure}
\ifx\pdftexversion\undefined
   \hskip 2cm \epsfig{file=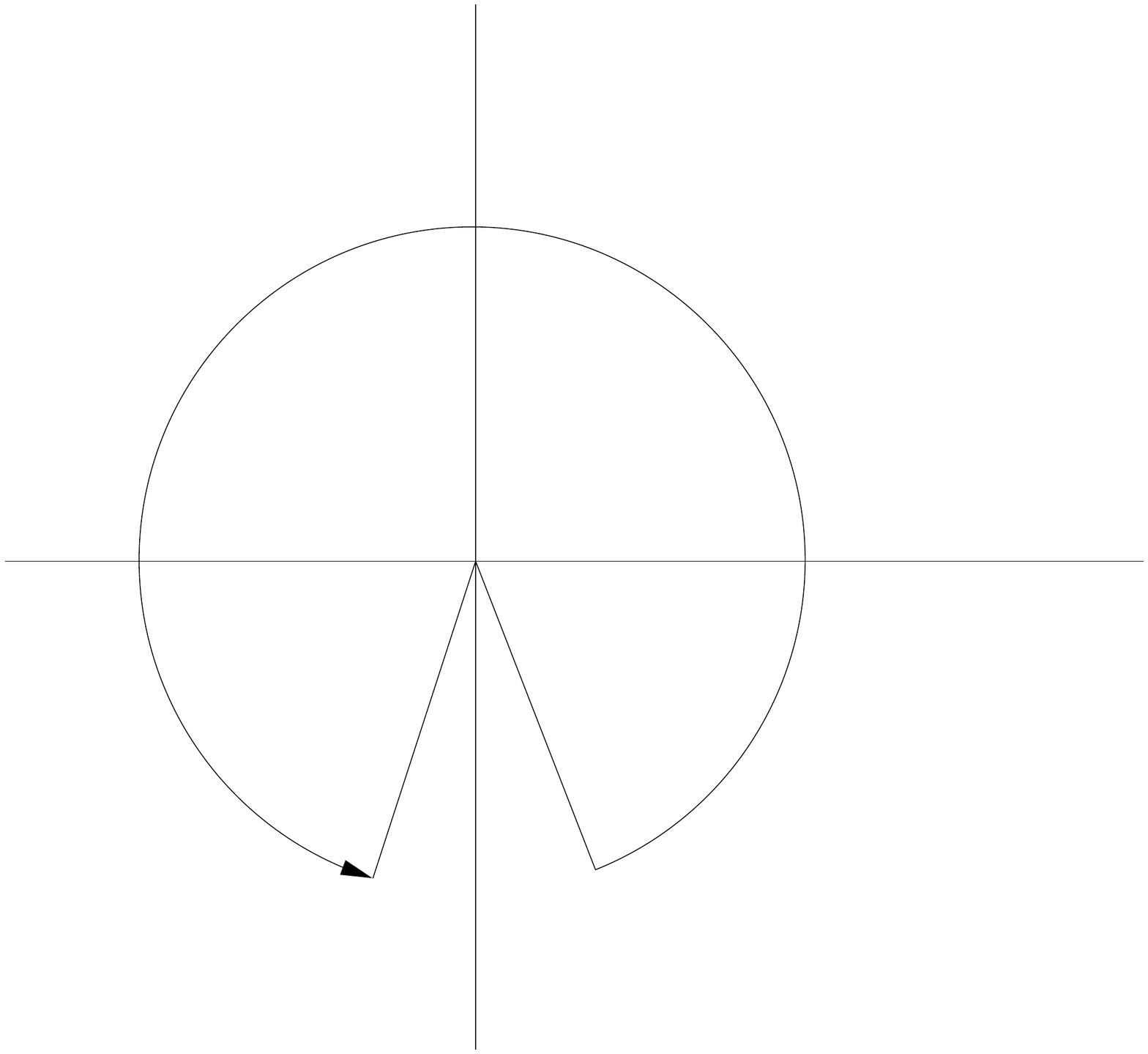, width=9cm,height=12cm}
\hskip0.5cm
\else
   \hskip 2cm \epsfig{file=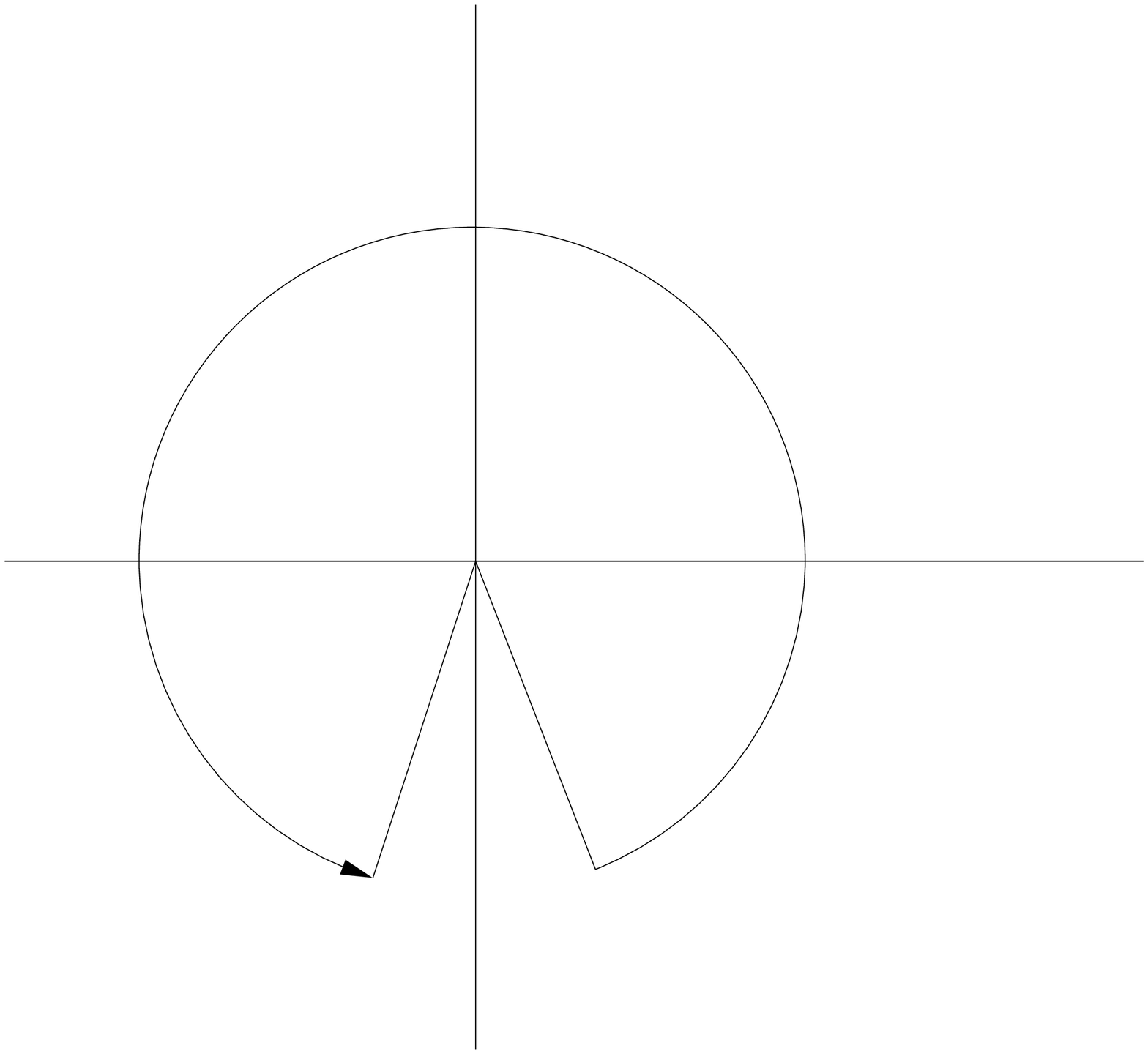, width=9cm,height=12cm}
\hskip0.5cm
\fi
\caption{The curve $\mathcal{C}$.}
\end{figure}
 Let 
\begin{equation}
  \label{defF}
  F(M)=\oint_{\mathcal{C}} \frac{e^{i\beta(s+s^{-1})}}{s^{M}}ds
\end{equation}
\begin{Proposition}\label{013}
We have 
\begin{equation}
  \label{mon2}
  \check{y}_{-1}(k)F(M+1)+\check{y}_{-2}(k)F(M+2)=0
\end{equation}\end{Proposition}
\z {\em Proof.} This follows immediately from the discussion above. $\Box$
\begin{Proposition}\label{014}
  For every large $N\in\NN$, $F(z)$ has exactly one zero of the form
  $z_N=N+o(N^0)$.  For large $N$ we have $F(1+z_{N})\ne 0$.
\end{Proposition}
\z {\em Proof.} It turns out that $F(M)$ is a Bessel function of order
$M$ evaluated at $2$ and a proof can be given based on this
representation.  However, in view of later generalizations we prefer
to give a more general argument that does not rely on explicit
representations.

Let $M=N+\zeta$ with $N$ a large positive integer, $\zeta$ complex
with $|\zeta|=\epsilon$ and $\epsilon$ positive and small. Let $C_1$ be the counterclockwise circle  $\{z:|z|=1\}$ and $L$ the segment $[0,-i]$; we write
\begin{multline}
  \label{dec2}
  F(N+\zeta)=(1-e^{-2\pi i\zeta})\int_L
\frac{e^{i\beta(s+s^{-1})}}{s^{M}}ds+\int_{C_1}
\frac{e^{i\beta(s+s^{-1})}}{s^{M}}ds\\=
2\pi i\zeta\Big(1+O(\zeta^{-1})\Big)\int_L
\frac{e^{i\beta(s+s^{-1})}}{s^{M}}ds+\int_{C_1}
\frac{e^{i\beta(s+s^{-1})}}{s^{M}}ds
\end{multline}
where in the integral along $L$ the principal branch of the log is
used.  The integral over $L$ can be estimated with Watson's Lemma, see
e.g.  \cite{benderorszag}
\begin{equation*}
 2\pi i\zeta\int_L
\frac{e^{i\beta(s+s^{-1})}}{s^{M}}ds=i\zeta (2\pi)^{3/2}\beta^{1-N} N^{N-3/2}e^{-N}e^{-\zeta\ln(\beta/N)}\Big(1+o(N^0)\Big)
\end{equation*}
By the Riemann-Lebesgue lemma, $\int_{C_1}\to 0$ as $N\to\infty$.  We
get
\begin{equation}
  \label{estF1}
  F(N+\zeta)= i^M\zeta (2\pi)^{3/2}\beta^{1-N} N^{N-3/2}e^{-N}e^{-\zeta\ln(\beta/N)}\Big(1+o(N^0,\zeta^0)\Big)+o(N^0)
\end{equation}
The existence of a unique simple zero at some $N+\zeta_N$ with
$|\zeta_N|<\epsilon$ is a consequence of the argument principle. 

\z The position of $\zeta_N$ can be found more accurately as follows. We
have
\begin{equation}
  \label{ratio}
  \zeta_N=\frac{\zeta_N}{1-e^{-2\pi i\zeta_N}}\int_{C_1}
\frac{e^{i\beta(s+s^{-1})}}{s^{N+\zeta_N}}ds\Big(\int_L
\frac{e^{i\beta(s+s^{-1})}}{s^{N+\zeta_N}}ds\Big)^{-1}
\end{equation}
from which it follows that $\zeta_N=o(\text{ \rm const}^N/N!)$ which readily implies that (\ref{ratio}) is contractive and that
\begin{equation}
  \label{ratio1}
  \zeta_N=\frac{1}{2\pi i}\int_{C_1}
\frac{e^{i\beta(s+s^{-1})}}{s^{N}}ds\Big(\int_L
\frac{e^{i\beta(s+s^{-1})}}{s^{N}}ds\Big)^{-1}\Big(1+o(N^0)\Big)
\end{equation}
Using (\ref{estF1}) and the fact that the first integral in
(\ref{ratio1}) gives the Laurent coefficients of
$e^{i\beta(s+s^{-1})}$ which can be independently estimated from the
series expansion, we find that, with constants that can be calculated,
\begin{equation}
  \label{estzetaN}
   |\zeta_N|=c_1 c_2^N N^{-2N+c_3}\Big(1+o(N^0)\Big)
\end{equation}
Thus $\zeta_{N+1}/\zeta_N\to 0$ as $N\to \infty$ and the second part
of the Proposition follows.
\begin{Proposition}
  Relation (\ref{mon2}), with $\check{y}_{-1}(k), \check{y}_{-2}(k)$ 
entire of exponential order one (cf. Remark \ref {R12}) implies 
\begin{equation}
  \label{concl1}
   \check{y}_{-1}(k)= \check{y}_{-2}(k)=0 \ \forall \, k\in\CC^3
\end{equation}
and then
\begin{equation}
  \label{concl12}
   {y}_{n}(x)= 0\  \forall  n\in\ZZ\  {\text{\rm and almost all}}\ x\in\RR^d
\end{equation}
\end{Proposition}

\z {\em Proof}. Propositions \ref{013} and \ref{014} and (\ref{defM})
imply that $\check{y}_{-1}(k)$ has at least $const\, R^2$ zeros in a
disk of large radius $R$.  Since $\check{y}_{-1}(k)$ is an entire
function of exponential order one, it follows that
$\check{y}_{-1}\equiv 0$ (see, e.g. \cite{SZ}). By (\ref{mon2}) we
have $\check{y}_{-2}\equiv 0$, so that ${y}_{-1}\equiv {y}_{-2}\equiv
0$ .

In the present model (\ref{hmdiff}) reads
\begin{equation}
  \label{hmdiffs}
  (-\Delta+\sigma+n\omega)y_n=-V_D\bchi_Dy_n-i\Omega_D\bchi_D(y_{n+1}-y_{n-1})
\end{equation}
and (\ref{hmdiffs}) with $n=-1$ and $n=-2$ implies $\bchi_D y_0=0$,
$\bchi_D y_{-3}=0$ respectively; inductively $\bchi_D y_n=0$ for all
$n$. Then $(-\Delta+\sigma+n\omega)y_n=0$ in $\RR^d\setminus\partial
D$. Since, by Proposition~\ref{Fl3}, $y_n\in L^2(\RR^3)$ we see (for
instance by taking the Fourier transform) that $\|y_n\|=0.$

\begin{Remark}\label{extension}
  It can be shown that condition {\bf (A)} holds with $V_D\bchi_D$
  replaced by $V_D\bchi_D+V_1(x)$ where $V_1(x)$ is bounded, not
  necessarily constant, with compact support disjoint from $D$. On the
  support of $V_1$, $\Omega$ is zero and it can be seen that for $n$
  sufficiently negative $y_n$ is zero on the support of $V_1$. From
  this point on the arguments are very similar, but we will not pursue
  this here.
\end{Remark}
\section{Compactness}\label{S6}
\subsection{Compactness of the operator $\mathfrak{g}_n$ defined in (\ref{GreenF2})} \label{compactness}  The case $d=1$ is discussed in Appendix~\ref{pd=1}. For $d=2,3$ compactness follows from Theorem VI.23 , Vol. 1, pp. 210
of \cite{Reed-Simon} (for $d=3$, note that
$e^{-\kappa_n|x-y|}/|x-y|\in L^2(D\times D)$).
\subsection{Compactness of $\mathfrak{C}$} 
The property (\ref{eq:Agmon}) is mentioned in Appendix A of
\cite{Agmon}. We include here an elementary proof of
(\ref{eq:normK}) below (which also can be refined without serious
difficulty to yield the sharper estimate (\ref{eq:Agmon})).
\begin{Lemma}\label{L22}
 We  have
\begin{equation}
  \label{eq:normK}
\|\mathfrak{g}_n\|{\longrightarrow} 0\ \ \ {\text{as}}\ |n|\to\infty
\end{equation}
(where $\|\cdot\|$ is the ${L^2(D)\mapsto L^2(D)}$ operator norm)
uniformly in $u$ in the region
$S_{\omega,\epsilon,A}=\{u:|u|<A,|\Re(u^2)|<\omega -\epsilon\}$, where
$A>0$ and $\epsilon$ is any small positive number.
\end{Lemma}
\z {\em Proof}.  Relation (\ref{eq:normK}) follows from a general
result by Agmon \cite{Agmon} which provides estimates on the rate of
convergence.  We give below an elementary proof in our case.

We prove the result for $d=3$ (the proof is simpler in $d=1,2$, noting
that for large $x$ with $\arg\,x \in (-\pi,\pi) $ we have
$K_{0}(x)=\sqrt{\frac{\pi}{2}}e^{-x}x^{-1/2}(1+o(1))$).  Define
\begin{equation}
  \label{defQ}
Q_{n}(x',x'')= \int_{{D}} dx
  \frac{e^{-\kappa_n|x'-x|-\overline{\kappa}_{n}|x''-x|}}{|x'-x||x''-x|}
\end{equation}
 We have
\begin{multline}\label{normK}
  \|\mathfrak{g}_n\|^2= \sup_{\|f\|=1}\int_{{D}^2}Q_{n}(x',x'')f(x')\overline{f(x'')}dx'dx''\\ \le \left(\int_{{D}^2}|Q_{n}(x',x'')|^2 dx'dx'\right)^{1/2}
\end{multline}
The last integral goes to zero as $|n|\to\infty$.  To see that, note
that
\begin{equation}
  \label{rel1}
\sqrt{n\omega+u^2}=\sqrt{n\omega}+O(n^{-1/2});\ \ {\text{as}}\   n\to+\infty
\end{equation}
and using the triangle inequality we get 
\begin{multline}
  \label{defQ2}
  \left|Q_{n}(x',x'')\right|\le {\rm
    Const\,}e^{-\sqrt{n\omega}|x'-x''|}\int_{{D}}
  \frac{dx}{|x'-x||x''-x|}\\\le {\rm
    Const\,}e^{-\sqrt{n\omega}|x'-x''|}
\end{multline}
and the conclusion, for $n\to +\infty$ follows by dominated
convergence.  We now focus on large negative $n$. Since
\begin{equation}
  \label{rel2}
\sqrt{n\omega+u^2}=-
i\sqrt{|n|\omega}+O(n^{-1/2});\ \ {\text{as}}\   n\to-\infty\end{equation}
it is easy to check that (\ref{eq:normK}) follows once we show
that $\|\mathfrak{g}_{[\nu]}\|\to 0$ as $\nu\to\infty$ where
$\|\mathfrak{g}_{[\nu]}\|$ is obtained by replacing $\kappa_n$ with
$i\nu$ in the definition of $\mathfrak{g}_n$.  We first show, with an analogous definition of $Q_{[\nu]}$, that
\begin{equation}
  \label{eq:boundQ}
\sup_{x,x'\in{D},\nu\in\RR}\left|Q_{[\nu]}(x,x')\right|=Q_0<\infty
\end{equation}
Indeed, we choose a ball $B_b$ centered at $x'$ of radius $b$ large
enough so that it contains ${D}$ and write the integrals (\ref{defQ})
in spherical coordinates centered at $x'$ with $x''$ on the $z$ axis;
in these coordinates $|x-x'|=r$ and $|x-x''|\ge d(x,Oz)=r\sin\theta$
and thus
\begin{equation}
  |Q_{[\nu]}(x,x')|\le \int_{B_b}\frac{dx}{|x-x'||x-x''|}\le
\int_{B_b}drd\theta d\phi\le 4\pi b
\end{equation}

Let $\rho(x;x',x'')=|x-x'|-|x-x''|$. We then have $|\rho(x;x',x'')|\le
|x'-x''|$ and we get
$$Q_{[\nu]}(x',x'')=\int_{{D}} dx
\frac{e^{i\nu\rho(x;x',x'')}}{|x'-x||x''-x|}=
\int_{-|x''-x'|}^{|x''-x'|}e^{i\nu\rho}d\mu(\rho)$$
where the positive
measure $\mu$ is defined by
\begin{equation}
  \label{eq:intgrd}
 \mu(A)= \mu_{x',x''}(A)=\int_{\{x:\rho(x)\in A\}\cap {D}}\frac{dx}{{|x'-x||x''-x|}}
\end{equation}
Since the integrand in (\ref{eq:intgrd}) is in $L^1$, the measure $\mu
$ is absolutely continuous with respect to the Lebesgue measure $m$.
We let $h(\rho;x',x'')=\frac{d\mu}{dm}$; then $h\in L^1$ and we get
\begin{equation}
  \label{Lebesgue}
Q_{[\nu]}(x',x'')  =
\int_{-|x''-x'|}^{|x''-x'|}e^{i\nu\rho}h(\rho;x',x'')d\rho
\end{equation}
By the Riemann-Lebesgue lemma we have\footnote{Noting that the
  estimate of the norm of $\mathfrak{g}_n$ can only increase if
  extended to $L^2(B)$ where $B$ is a ball containing $D$, and that
  $h$ calculated in $B$ is piecewise smooth the max of $Q_{[\nu]}$ can
  be in fact bounded by an inverse power of $\nu$; we do not however
  need this refinement here.}
\begin{equation}
  \label{eq:Qtozero}
Q_{[\nu]}(x',x'') \to 0\ \text{as}\ \nu\to\infty  
\end{equation}
Now (\ref{eq:boundQ}), (\ref{eq:Qtozero}) and again dominated
convergence implies $ \|\mathfrak{g}_{[\nu]}\|\to 0$ as $\nu\to\infty$
completing the proof.
\begin{Lemma}\label{L27}
  Under the assumption (\ref{compactDa}), the operator $\mathfrak{C}$
  is compact on $\mathcal{H}$ and analytic in $u$ in
  $S_{\omega,\epsilon, A}$, cf. Lemma~\ref{L22}.
\end{Lemma}
\z Indeed, $\mathfrak{C}$ is the norm limit (\ref{limC}), uniform in
$u\in S_{\omega,\epsilon, A}$, where $\mathfrak{C}_N$ are compact by
Lemma~\ref{L22} and analytic as explained in the proof of
Lemma~\ref{L6}. We note that the operator 
\begin{equation}
  \label{normOmega}
 \left\|\sum_{j\in\ZZ}\Omega_j S^{-j}\right\|
\end{equation}
is bounded in $\mathcal{H}$. Indeed, if we write $\langle
n\rangle:=1+|n|$ we have, for $(n,j)\in\ZZ^2$ $\langle n\rangle \le
\langle j\rangle\langle n-j\rangle$ and
\begin{multline}\label{Yajima1}
 \sum_{n\in\ZZ}\langle n\rangle^{\gamma}\left|\sum_{j\in\ZZ}\Omega_j y_{n-j}\right|^2\le
\sum_{n\in\ZZ}\left(\sum_{j\in\ZZ} \langle j\rangle^{\gamma/2}|\Omega_j| \langle n-j\rangle^{\gamma/2}|y_{n-j}|\right)^2 \\ = 
\sum_{j_1,j_2\in\ZZ}\langle j_1\rangle^{\frac{\gamma}{2}} 
|\Omega_{j_1}|
\langle j_2\rangle^{\frac{\gamma}{2}}
 |\Omega_{j_2}|
\sum_{n\in\ZZ}
\langle n-j_1\rangle^{\frac{\gamma}{2}}|y_{n-j_1}|
\langle n-j_2\rangle^{\frac{\gamma}{2}}|y_{n-j_2}|\\
\le \|y\|_{l^2_\gamma}^2\|\langle j\rangle^{\gamma/2}\Omega_j\|^2_1\le C\|y\|_{l^2_\gamma}^2
\end{multline}
by (\ref{Fourier}).
$\Box$
\section{Appendixes}
\def\thesubsection{\Alph{subsection}}
\subsection{Compact operator formulation and proof of Lemma~\ref{L4} for  $d=1$}\label{pd=1}  We can assume without loss of generality $D\subset[-1,1]$. For $n=0$ we choose some large
positive $a$ such that $\sin 2\sqrt{a}\ne 0$, denote by $f_{\pm}(x)$
the functions $e^{\mp ux}$ and let $\psi_+$ be  the solution of the
equation
\begin{equation}
  \label{d=1,1}
  -\psi''+(a \bchi_{[-1,1]}+ u^2)\psi=0
\end{equation}
(see Remark~\ref{R4}) with initial condition $\psi_+(1)=f_+(1)$,
$\psi'_+(1)=f'_+(1)$, and similarly let $\psi_-$ be the solution of
(\ref{d=1,1}) with initial condition $\psi_-(-1)=f_-(-1)$,
$\psi'_-(-1)=f'_-(-1)$. Since both the equation and the initial
conditions are analytic in $u$ at $u=0$, so are the solutions
$\psi_{\pm}$ and their Wronskian $W(u)$.  It can be checked that
$W(0)=\sqrt{a}\sin 2\sqrt{a}\ne 0$. In fact, taking
$\tau_a=\sqrt{a-u^2}$ we have
\begin{equation}
  \label{pm}
  \psi_{\pm}(x)=\tau_a^{-1}e^{-u}\Big[\tau_a\cos(\tau_a x\mp \tau_a)\mp u\sin(\tau_a x\mp \tau_a))\Big]
\end{equation}
In a neighborhood of $u=0$ we write for $n\ne 0$ the same integral
expression (\ref{eq:fint2}), while for $n=0$ we write 
\begin{equation}
  \label{d1,2}
 y_0=\mathfrak{g}_{0,a}\psi_{1,0}-\mathfrak{g}_{0,a} (V +a \bchi_{[-1,1]}) y_0+i\mathfrak{g}_{0,a}\left(\sum_{j\in\ZZ}\Omega_j S^{-j}y\right)_n\end{equation}
where 
\begin{equation}
  \label{eq:defg0}
  W(u)(\mathfrak{g}_{0,a} f)(x)=\psi_{+}(x)\int_{-1}^x \psi_{+}(s)f(s)ds-
\psi_{-}(x)\int_{1}^x \psi_{-}(s)f(s)ds
\end{equation}
With the same conventions, we now write the integral system in the
form (\ref{eq:fint2}). Compactness and analyticity are now shown in
the same way as for $d=2,3$. $\Box$.
\subsection{Proof of Lemma~\ref{L5}}\label{Unic} Let $\sigma=\sigma_0-2i\tau$ where $\sigma_0\in[0,\omega)$ and $\tau>0$. We show that 
\begin{equation}
  \label{normc1}
  \|\mathfrak{C}\|\to 0\ \  {\text as }\ \  \tau\to\infty
\end{equation}
and uniqueness follows by contractivity. The calculation leading to
(\ref{normc1}) is quite straightforward, but we provide it for
convenience.  In $d=1,2$ the estimate follows from the behavior of the
Green function for large argument. We then focus on $d=3$. By
(\ref{defkappa}) we have
\begin{equation}
  \label{eq:evalroot}
  \Re(\kappa_n)=\left(\frac{1}{2}\left((\sigma_0+n\omega)^2+\sigma_0+n\omega\right)^{1/2}+\tau\right)^{1/2}
\end{equation}
For $n>0$ we then have $\Re\kappa_n>\sqrt{n\omega}$ and the same calculation
as for (\ref{defQ2}) shows that
\begin{equation}
  \label{eq:normK1}
\|\mathfrak{g}_n\|\mathop{\longrightarrow}_{L^2(B)}  0\ \ \ {\text{as}}\ n\to +\infty
\end{equation}
uniformly in $\tau$.
For $n<0$ (\ref{eq:evalroot}) gives
\begin{equation}
  \label{Qn23}
  \left|Q_{n}(x',x'')\right|\le \left|Q_{\nu}(x',x'')\right|
\end{equation}
where 
\begin{equation}
  \label{eq:evalroot3}
  -\nu:=\Im(\kappa_n)\to\infty \ \text{as } n\to-\infty
\end{equation}
and now (\ref{eq:Qtozero}) shows that 
\begin{equation}
  \label{eq:normK2}
\|\mathfrak{g}_n\|\mathop{\longrightarrow}_{L^2(B)}  0\ \ \ {\text{as}}\ n\to -\infty
\end{equation}
uniformly in $\tau$. We choose then $n_0$ large enough so that
\begin{equation}
  \label{eq:normK23}
\sup_{n\ge n_0,\tau>0}\|\mathfrak{g}_n\|\le \epsilon
\end{equation}
For $\tau$ large enough we have, still from (\ref{eq:evalroot}),
\begin{equation}
  \label{midr}
  \Re(\kappa_n)>\frac{1}{2}\tau^{1/2};\ \ -n_0\le n\le n_0
\end{equation}
Choosing a ball $B$ centered at $x$ containing $D$, we then have for
large $\tau$ and some constants independent of $f, \tau,x$ and
$n\in(-n_0,n_0)$, with the notation $\alpha=\frac{1}{2}\tau^{1/2}$,
\begin{multline}
  \label{eq:partK2} \left|\Big(\mathfrak{g}_{n} f\Big)(x)\right|\le
  \left\|\frac{e^{-\alpha|x-x'|}}{|x-x'|}\right\|_{L^2(D)}\|f\|_{L^2(D)} \\ \le
  C_1\|f\|_{L^2(D)}\left\|\frac{e^{-\alpha|x-x'|}}{|x-x'|}\right\|_B\le
\frac{C_2}{\tau}\|f\|_{L^2(D)}\le\epsilon
\end{multline}
and the conclusion follows.
\subsection{Proof of Lemma~\ref{L12}}\label{simpoles} 
Assume $\sigma_0$ is a value of $\sigma$ where invertibility of
$I-\mathfrak{C}(\sigma_0)$ fails. Then $\sigma_0\in[0,\omega)$. By the
Fredholm alternative we know that $I-\mathfrak{C}(\sigma)$ is
invertible in some punctured neighborhood of $\sigma_0$ where the
solution of (\ref{eq:fint2}) is meromorphic.

(i): $\sigma_0\ne 0$.  Denote $\zeta=\sigma-\sigma_0$. We rewrite
(\ref{eq:fint2}) in a suitable way near $\sigma_0$. We have from
(\ref{eq:finy})
\begin{equation}
  \label{mod1}
  (-\Delta+\sigma_0+n\omega)y_n=-i\psi_{2,n}-Vy_n-\zeta y_n+
\sum_{j\in\ZZ}\Omega_j(x)\left(S^{-j}y\right)_n
\end{equation}
which we write symbolically 
\begin{equation}
  \label{mod12}
  \mathfrak{W}y=-\zeta y-i\psi_{2,n}
\end{equation}
and from (\ref{eq:finy}) and (\ref{eq:fint2}) we have *************
\begin{equation}
  \label{eq:fint3}
 y_n=-i\mathfrak{g}_n\psi_{2,n}-\mathfrak{g}_n \Big[V y_n-\zeta y_n-\sum_{j\in\ZZ}\Omega_j \left(S^{-j}y\right)_n\Big]
\end{equation}
implying the following version of (\ref{eq:fint2}), with evident
notation,
\begin{equation}
  \label{it2}
  y=y_0+\zeta\mathfrak{g}y +\mathfrak{C}(\sigma_0)y
\end{equation}
On the other hand,
\begin{equation}
  \label{laurent}
  y=\sum_{j=-M}^{\infty}c_j\zeta^j
\end{equation}
with the coefficients $c_j\in\mathcal{H}$. Assume, to get a
contradiction, that $M\ge 2$. Inserting in (\ref{it2}) we get
\begin{eqnarray}
  \label{f0}
  c_{-M}&=&\mathfrak{C}(\sigma_0) c_{-M}\\
 c_{-M+1}&=&\mathfrak{C}(\sigma_0) c_{-M+1}+\mathfrak{g}c_{-M}\nonumber\\
\cdots\nonumber
\end{eqnarray}
In differential form we have,
\begin{eqnarray}
  \label{f01}
  \mathfrak{W}c_{-M}&=&0\\
 \mathfrak{W}c_{-M+1}&=&-c_{-M}
\end{eqnarray}
By (\ref{f0}) and Proposition~\ref{Fl3} we have $c_{-M}\in
l^2(L^2(\RR^d)):=\mathcal{H}_1$ (in fact, $(c_{-M})_n$ decay at least
exponentially in $|x|$). On the other hand we then have from
(\ref{f01}) and noting the formal self-adjointness of $\mathfrak{W}$,
\begin{equation}
  \label{mod3}\langle c_{-M}, c_{-M}\rangle=
 -\Big\langle c_{-M},\mathfrak{W}c_{-M+1}\Big\rangle=
-\Big\langle\mathfrak{W} c_{-M},c_{-M+1}\Big\rangle=0
\end{equation}
which is a contradiction. 

(ii) $\sigma_0=0$: there are two differences w.r.t case (i): (a)
meromorphicity and Laurent expansions now use the variable
$u=\sqrt{\sigma}$; and (b) $c_{-M}$ is not necessarily in
$\mathcal{H}_1$ so we work with $\mathcal{H}_B=l^2(L^2(B))$ for large
enough $B$. These differences can be dealt with straightforwardly, so
we just outline the main steps.  The Laplacian is the only ingredient
of $\mathfrak{W}$ not formally self-adjoint in $\mathcal{H}_B$.
Integration by parts, implicit in (\ref{mod3}) brings in boundary
terms of the form
\begin{equation}
  \label{eq:Bt}
  \int_{\partial B}f\nabla g\cdot dS
\end{equation}
where $f$ and $g$ are $c_{-M}$ or $c_{-M+1}$. Both $f$ and $g$ have
decay $|x|^{-1}$ and this behavior is differentiable, as is manifest
from (\ref{f0}), (\ref{formulares}), and (\ref{GreenF2}).  The
contribution from the integrals (\ref{eq:Bt}) is thus $O(r_B^{-1})$
which equals the norm $\|c_{-M}\|_{\mathcal{H}_B}$, clearly
nondecreasing in $r_B$. This again forces $c_{-M}=0$, a contradiction.
$\Box$
\subsection{Sketch of the proof of Proposition~\ref{genth1} and Theorem~\ref{genth12}}\label{Sketch} We first show Theorem~\ref{genth12} (i). 
The contour of the inverse Laplace transform can be deformed as shown
in Fig. 2. Pushing the contour of integration to the left brings in
residues due to the meromorphic integrand, and since the kernel of the
inverse Laplace transform is $e^{pt}$, residues in the left half plane
give rise to decaying exponentials in $\psi(x,t)$. Uniform bounds on
the Green function as $p\to -\infty$ are easy to prove. Consequently,
there are only finitely many arrays of poles of $\hat{\psi}$. The
contour of integration in the inverse Laplace transform can be pushed
all the way to $-\infty$ in view of the exponential decay of the
kernel $e^{pt}$.  We are left with integrals along the sides of the
cuts which, after the change of variable $p\leftrightarrow -p$ (or
$p\leftrightarrow -pe^{i\alpha}$ if poles exist on the cuts), are seen
to be Laplace transforms.  Since $\hat{\psi}$ is analytic in
$\sqrt{p+i n\omega}$, the contour deformation result shows, ipso
facto, Borel summability of the asymptotic series of $\psi(x,t)$ for
large $t$.

The general case is proved in a very similar way, using
Lemma~\ref{L12}. If {\bf (A)} does not hold, then some of the poles of
the meromorphic function $y_n(\sigma)$ can be on the segment
$(-\omega,\omega)$. If a pole is placed at $\sigma=0$, then
analyticity in $u$ in the operator entails a singularity of the form
$\sigma^{-1/2}A(\sigma)+B(\sigma)$ with $A$ and $B$ analytic, whence
the conclusion.

\begin{figure}
\ifx\pdftexversion\undefined
   \hskip 2cm \epsfig{file=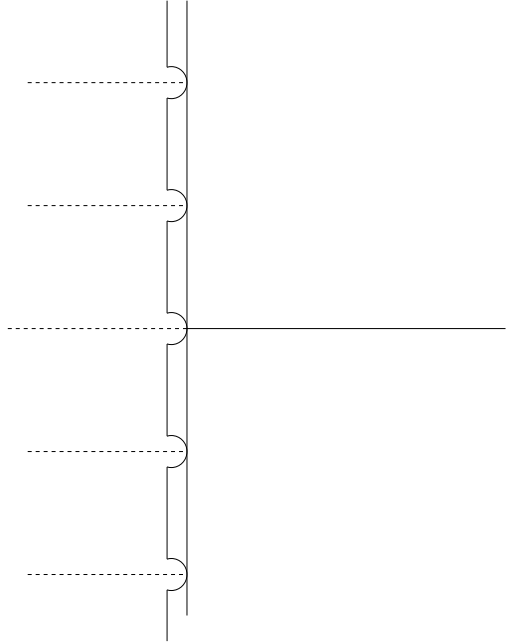, width=9cm,height=12cm}
\hskip0.5cm
\else
\hskip 2cm \epsfig{file=fig1.ps, width=10cm,height=12cm}
\hskip0.5cm
\fi
\caption{Deformation of the integration contour.}
\end{figure}
\subsection{Estimates needed for Proposition~\ref{P17} and  Proposition~\ref{Fl3}}\label{PP21} 
\begin{Proposition}\label{P161}
  Let $y$ be a solution in $l^2(L^2(B))$ of the homogeneous equation
  associated to (\ref{eq:finy}).  Under the assumptions
  (\ref{compactDb}), we have
\begin{equation}
  \label{1/j}
\|  y_j\|_{L^2(B)}=O(j^{-2})\ {\text as } |j|\to\infty
\end{equation} 

\end{Proposition}
Let $\epsilon$ be small enough and choose $j_0>0$ large enough (the
proof is similar for $j_0<0$) so that
$\|\mathfrak{g}_j\|_{L^2(B)}<\epsilon$ for all $j\ge j_0$, see
(\ref{eq:normK}). We consider the Banach space $\mathcal{B}_{j_0}$ of
sequences of functions $\{y_j\}_{j\ge j_0}$ defined on $B$ for which
the norm
  \begin{equation}
    \label{NormB}
  \|y\|_{j_0}:=  \sup_{j\ge j_0}j^{2}\|  y_j\|_{L^2(B)}
  \end{equation}
  is finite. For $j>j_0$ we write the homogeneous part of
  (\ref{eq:finy}) in the form
\begin{multline}
  \label{eq:fintq}
  y_j=-\mathfrak{g}_j V y_j+i\mathfrak{g}_j\Big[\sum_{m\ge 0}\Omega_{-m}
  y_{m+j}(x) \\+\sum_{0\le l \le j-j_0}\Omega_l y_{j-l}(x)+\sum_{l\ge
    j-j_0}\Omega_l y_{j-l}(x)\Big]
\\=-\mathfrak{g}_j V y_j+i\mathfrak{g}_j\Big[\sum_{m\ge 0}\Omega_{-m}
  y_{m+j}(x)+\sum_{0\le l \le j-j_0}\Omega_l y_{j-l}(x)\Big] +E_j(x)\\=:
\Big(\mathfrak{T}_{j_0}\,\,y+E\Big)_j
\end{multline}
Since $\|\Omega_j\|_{L^2(B)}=O(j^{-2})$ and $y\in l^2(L^2(B))$ we see
that $\|E\|_{j_0}<\infty$. It can be checked that
$\mathfrak{T}_{j_0}:\mathcal{B}_{j_0}\to \mathcal{B}_{j_0}$ is
bounded, that $\|\mathfrak{T}_{j_0}\|\to 0$ as $j_0\to \infty$, and
thus eq.  (\ref{eq:fintq}) is contractive if $j_0$ is large. The
Proposition follows.

\z {\bf Acknowledgments}.  We are very grateful to K. Yajima for many
useful comments and suggestions, including the argument in
(\ref{Yajima1}).  We thank A. Soffer for helpful discussions.  Work
supported by NSF Grants DMS-0100495, DMS-0074924, DMR-9813268, and
AFOSR Grant F49620-01-1-0154.

\end{document}